\documentclass[twocolumn,floatfix]{revtex4}

	\usepackage{amssymb}
	\usepackage{amsmath}
	\usepackage{epsfig}
	\usepackage{graphicx}

\begin{document}
\title{Persistence and periodicity in a dynamic proximity network\footnote{Published in the proceedings of the DIMACS Workshop on Computational Methods for Dynamic Interaction Networks (Piscataway), 2007.}}
\author{Aaron Clauset${}^{\ddag,*}$ and Nathan Eagle${}^{\dag}$}
\affiliation{\mbox{${}^{\ddag}$Santa Fe Institute, 1399 Hyde Park Rd., Santa Fe NM 87501}} 
\affiliation{\mbox{${}^{*}$Department of Computer Science,
University of New Mexico, Albuquerque NM 87131} \\
{\tt aaronc@santafe.edu} \\
\mbox{${}^{\dag}$Media Laboratory,
Massachusetts Institute of Technology, Cambridge, MA 02139} \\
{\tt nathan@media.mit.edu} }

% ---------- Abstract / Opener ----------
\begin{abstract}
The topology of social networks can be understood as being inherently dynamic, with edges having a distinct position in time. Most characterizations of dynamic networks discretize time by converting temporal information into a sequence of network ``snapshots'' for further analysis. Here we study a highly resolved data set of a dynamic proximity network of 66 individuals. We show that the topology of this network evolves over a very broad distribution of time scales, that its behavior is characterized by strong periodicities driven by external calendar cycles, and that the conversion of inherently continuous-time data into a sequence of snapshots can produce highly biased estimates of network structure. We suggest that dynamic social networks exhibit a natural time scale $\Delta_{\rm nat}$, and that the best conversion of such dynamic data to a discrete sequence of networks is done at this natural rate.
\end{abstract}

\maketitle

% ---------- Introduction ----------
\noindent Complex systems of interacting components are now often represented as a network, i.e., $n$  nodes or vertices joined together in pairs by $m$ links or edges. Characterizing these networks' topological patterns can often yield significant insights into the structure and function of the original system~\cite{Newman03a, Barabasi02, Dorogovtsev02}, and networks from a wide variety of domains, including social~\cite{Watts98, Scott00, Newman03b}, technological~\cite{Albert99, Kleinberg01, Clauset05} and biological~\cite{Williams00, Jeong00, ShenOrr02} systems, have been studied in this way. From this kind of network analysis, it has been demonstrated that many real-world networks exhibit similar properties. For instance, most real-world networks exhibit a highly heterogeneous degree distribution and short topological distances between arbitrary vertices.

Social networks differ from many other kinds of networks~\cite{Newman03c}, for instance, by having a larger-than-expected density of short loops (typically triangles -- the so-called ``clustering coefficient'' although the behavior seems to generalize to short loops of all lengths). Further, social networks exhibit a pattern in connectivity at the whole-network level that is often called ``community structure,'' in which large and relatively dense subgraphs are themselves sparsely connected together~\cite{Newman02a}, as well as assortative mixing on various node attributes~\cite{Newman02}.

\begin{figure} [b]%[hbp]
\begin{center}
\includegraphics[scale=0.45]{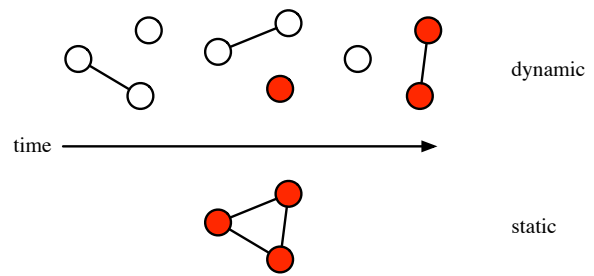}
\caption{A schematic illustrating the impact of the temporal ordering of social interactions for an epidemic process (colored nodes) spreading on the network. In contrast to the lower panel, in which all three nodes appear to be connected, the upper panel shows that interactions that precede the time of infection do not propagate the contagion. }
\label{fig:ordering}
\end{center}
\end{figure}

Much of our understanding of the large-scale structural patterns in social networks has come from the study of static topologies -- an idealization that naturally omits any dynamic or time-evolving character of the social processes that underly these networks. For some scientific questions, a static topology may be sufficient. However, for questions regarding dynamics, the temporal variation of edges themselves is likely to be important. For instance, in the spread of an epidemic, the order of interaction can have a significant effect on which individuals become infected, and ultimately the size of an epidemic outbreak, e.g., if two individuals $A$ and $B$ interact prior to $A$ being infected, but not after, then $B$ has no risk of infection, a dynamic that is difficult to capture with static topologies (Fig.~\ref{fig:ordering}).

Unfortunately, temporal connectivity data is often difficult to obtain for real social systems. The traditional approach to studying network dynamics in sociology (see, for instance,~\cite{wasserman:faust:1994} and reference therein) tends to rely heavily on interaction data self-reported by study participants, which exhibit significant bias and noise~\cite{Marsden90}. Several recent studies in physics and computer science have utilized web-based or other indirect sources of dynamic social network data~\cite{Onody04,Holme03,Holme04a, Vazquez05, Kleinberg05,Saia06}. In general, empirical studies convert the available temporal data into a short sequence of non-overlapping network ``snapshots,'' (called ``panel data'' in the sociology literature~\cite{Snijders06}), each of a length of time greater than the natural time scale of topological variation. This sequence can then be further analyzed to extract its time-varying structure, e.g., by converting the networks into a scalar time series of some network statistic, or computing statistics over the entire sequence. A notable exception to this emphasis on discrete time-series analysis of dynamic behavior is the work of Kempe, Kleinberg and Kumar~\cite{Kempe00} who consider theoretical questions related to processes on networks with edges that vary in continuous time.

In this paper, we study the structural patterns of a dynamic social network of the sort considered by Kempe, Kleinberg and Kumar, that is, a network whose edges effectively vary continuously in time. Wed draw our data from the time-varying physical proximity of 115 individuals over the course of one month in the MIT Reality Mining study~\cite{Reality}. Here, we consider several questions about the temporal connectivity patterns in this data. First, we show that the persistence of proximity in this data set appears to consistently follow a heavy-tailed distribution -- possibly having log-normal form -- an observation that is, to our knowledge, novel. We then consider to what degree these proximity networks exhibit periodicity, driven by the daily and weekly rhythms of human social behavior. Finally, we consider the question of whether there is a natural length of time over which to aggregate edges such we preserve the important temporal variation, while smoothing out some of the high-frequency variation in the data.

\vspace{3mm}\noindent {\bf \large Network Statistics}

Before we begin our analysis, we will first briefly review a few commonly used network metrics for static topologies~\cite{Newman03a}, and their extension to dynamic topologies. We then introduce a similarity statistic to measure the degree of topological overlap between two networks, e.g., between two sequential snapshots.

Recall that the adjacency matrix $A$ of a simple graph is an $n\times n$ $(0,1)$-matrix with a zero diagonal, where $n$ is the number of nodes in the network, and an element $A_{ij}=1$ if and only if vertices $i$ and $j$ are connected, and $A_{ij}=0$ otherwise. Physical proximity is inherently an undirected quantity and thus our adjacency matrices are symmetric. The degree $k$ of a vertex $i$ is thus simply the row-sum $\sum_{j=1}^{n}A_{ij}$.

Our proximity data appear as tuples of the form $(i,j,t_{1},t_{2})$, denoting that $i$ and $j$ are proximate to each other starting at time $t_{1}$ and ending at time $t_{2}$, where $t_{1}$ and $t_{2}$ are effectively real-valued numbers~\cite{footnote1}. In order to transform this information into a sequence of $T$ network snapshots $\vec{A} = A^{(1)},A^{(2)},\dots,A^{(T)}$, we must first choose a length of time $\Delta$, which we call the {\em snapshot rate}, that each snapshot covers. We then simply say that $A^{(t)}_{ij}=1$ if and only if the nodes $i$ and $j$ are proximate at any time between $t$ and $t+\Delta$, and $A_{ij}=1$ otherwise. A natural question, which we will consider shortly, is what effect the free parameter $\Delta$ has on the observed patterns in dynamic network topology. When both $\Delta$ and the density of nodes in physical space are small, these snapshots will necessarily be very sparse, each containing only a few edges. In this sense, these snapshots differ markedly from most other social networks in which the average degree $\langle k \rangle > 1$ and a large connected component exists. 

In addition to calculating two standard statistical measures of network structure -- the average degree $\langle k \rangle$ of a node and the local density of triangles $C$ (also called the clustering coefficient~\cite{footnote2}) -- over $\vec{A}$, we also calculate a similarity measure of a vertex's connectivity at one snapshot to the next, which we call the {\em adjacency correlation} $\gamma$.

Given a pair of adjacency matrices $A^{(x)}$ and $A^{(y)}$, one definition of the similarity of the connectivity for a vertex $j$ at time $x$ and time $y$ would be the Pearson correlation coefficient on the $j$th row vector at time $x$ and at time $y$. However, in a sparse graph, most entries in each row are zero and our measure of similarity would be uninformatively large. A more useful measure would compute the correlation between the row elements that are non-zero at least once in either of the two matrices; letting $N(j)$ denote this union, the adjacency correlation for $j$ becomes
\[ \gamma_{j} = \frac{\sum_{i\in N(j)} A^{(x)}_{i,j}~A^{(y)}_{i,j} }{\sqrt{ (\sum_{i\in N(j)} A^{(x)}_{i,j})~(\sum_{i\in N(j)} A^{(y)}_{i,j} )}} \enspace .
\]
When the adjacency vectors contain no edges, the adjacency correlation is undefined, so for convenience, we take $\gamma_{j}=1$ in this case. Averaging $\gamma$ over all vertices yields a statistic that represents the average topological overlap of the neighbor set between two snapshots.

\begin{figure} [t]%[hbp]
\begin{center}
\includegraphics[scale=0.45]{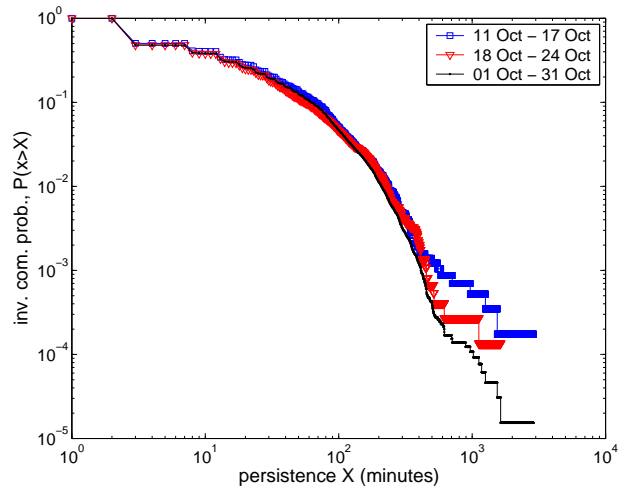}
\caption{The empirical distribution of the persistence of edges in the network on doubly logarithmic axes, for two week during and the full month of October for the core 66 subjects. Clearly, the network topology is evolving over a broad range of time-scales, with both fast (less than $10$ minute) and slow (greater than 100 minute) variations. Surprisingly, the distributions appear to be roughly identical except for the upper tail.  }
\label{fig:persistence}
\end{center}
\end{figure}

\begin{figure*}[t] %[tpbh]
\begin{center}
\begin{tabular*}{17.5cm}{lll}
\includegraphics[scale=0.333]{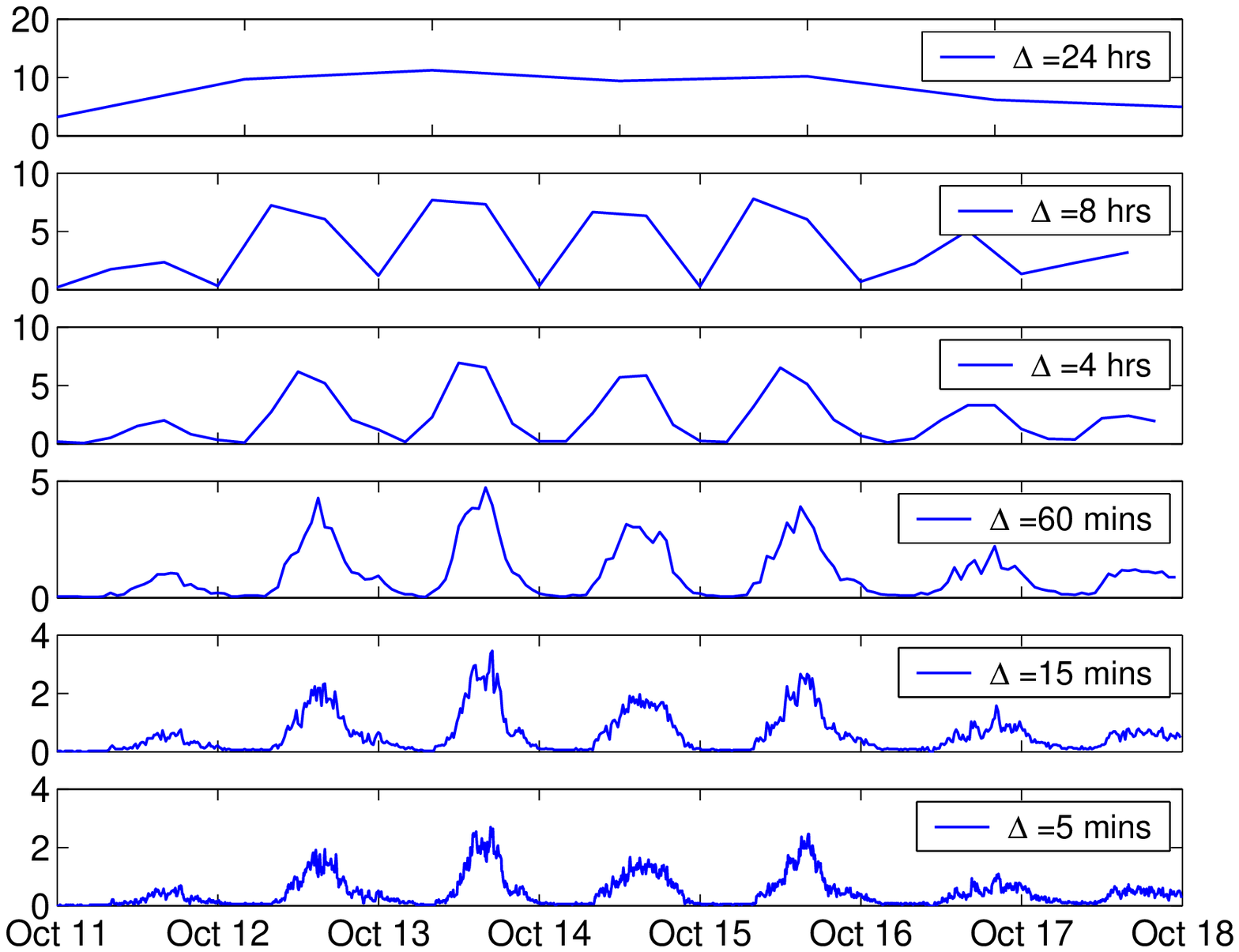} &
\includegraphics[scale=0.333]{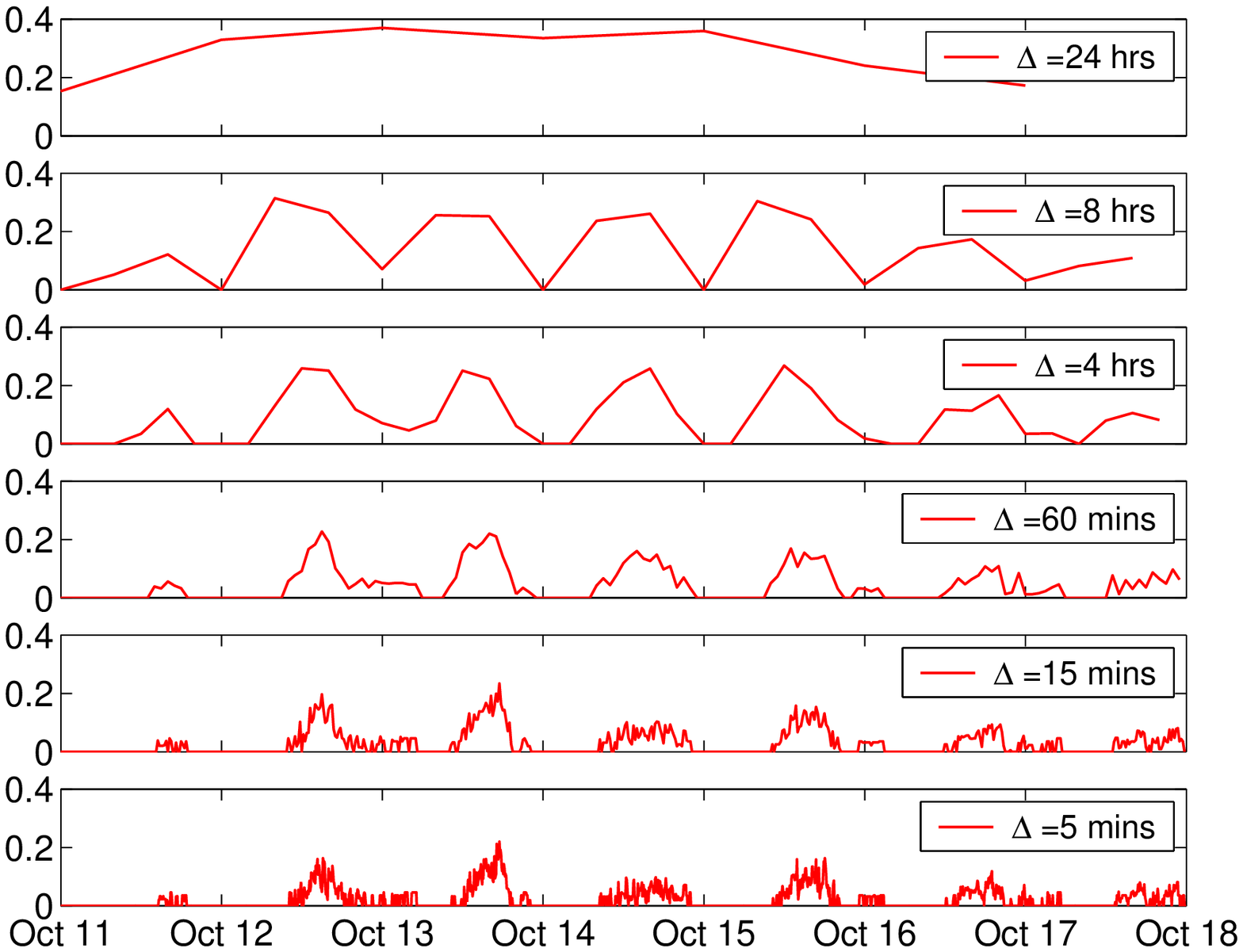} &
\includegraphics[scale=0.333]{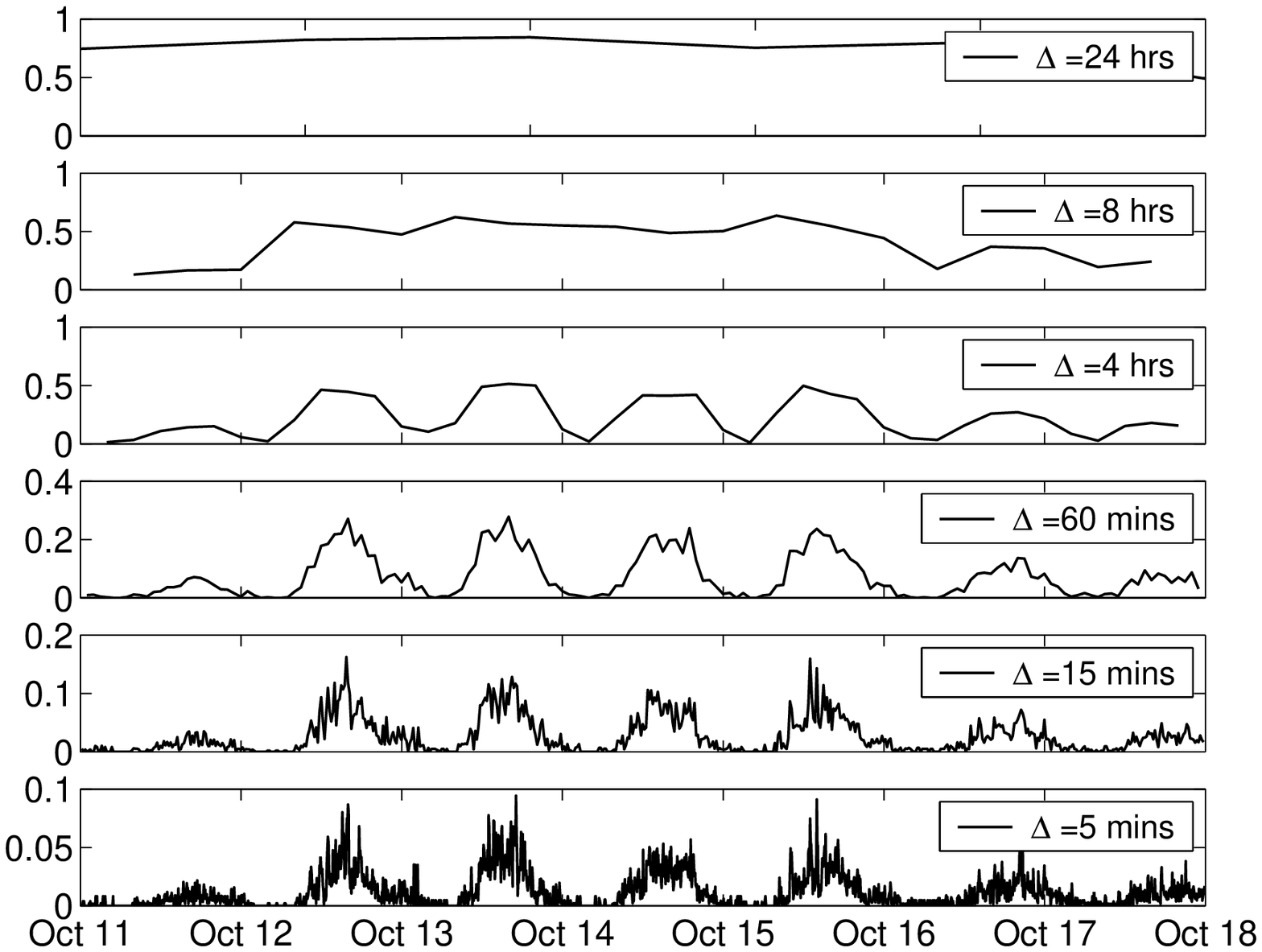} \\
\end{tabular*}
\end{center}
\caption{The (a) mean degree $\langle k\rangle$, (b) mean clustering coefficient $C$ and (c) the network adjacency correlation $1-\gamma$ as a function of time for snapshot rates $\Delta = \{1440, 480, 240, 60, 15, 5\}$ (minutes) during the week of  11 October through 17 October for the core 66 subjects. As $\Delta$ grows, under-sampling clearly averages out higher frequency fluctuations.}
\label{fig:timeseries}
\end{figure*}

\vspace{3mm}\noindent {\bf \large Network Analysis}

We now turn to the question of what impact altering the snapshot rate $\Delta$ has on the observed dynamic topology has on the aforementioned structural statistics.

The Reality Mining study followed 115 subjects at MIT carry proximity-aware mobile phones for roughly nine months, collecting over $500~000$ hours of proximity data~\cite{Reality}. At five minute intervals, each phone completes a Bluetooth device discovery scan over its local area (7-10m diameter) and records the identities of all discovered devices, including the other subjects' phones. From this data, we first extract adjacencies for each scan, and then infer ongoing proximity so as to annotate each edge with an initiation and termination time. Here, we restrict our analysis to the $24~092$ undirected subject-subject adjacencies between 01 October 2004 and 31 October 2004 of the 66 subjects who work in the same building at MIT.

From the starting and ending times of each edge, we compute its temporal duration, or persistence, as $t_{2}-t_{1}$. Figure~\ref{fig:persistence} shows the empirical distribution function for these durations (in minutes) for both two weeks within and the full month of October. Unsurprisingly, the weekly and monthly distributions are largely similar with minor variations in the upper tail ($x>400$ minutes), suggesting a system largely in equilibrium, or with strongly regular features. The average persistence of an adjacency is relatively small ($\langle x \rangle =  22.83$ minutes) while there are several (three) edges that persist for more than 1440 minutes (24 hours). The large variations in persistence demonstrate that the network's topology evolves at a broad range of time-scales. We conjecture that much of this regularity is driven by the strong periodicities in human behavior, e.g., the home-work-home daily cycle, and the work-home-work weekly cycle.

Now we turn to the question of what impact the snapshot rate $\Delta$ may have on the observed network dynamics. We note that choosing a $\Delta$ larger than the natural time-scale of the topological variation of the network will naturally cause high-frequency variations to be averaged out, potentially obscuring important variation in ordering such that depicted in Figure~\ref{fig:ordering}. In most empirical studies of dynamic networks, the snapshot rate is often determined by the method of data collection, or chosen to guarantee a certain density of edges in each snapshot; here, we use our highly temporally resolved data to explore the question of what kind of artifacts can be introduced by an incorrect choice of $\Delta$, and whether there might be a natural choice for proximity networks.

Choosing snapshot rates \mbox{$\Delta = \{1440, 480, 240, 60, 15, 5\}$} minutes, we compute the mean degree $\langle k\rangle$ (Fig.~\ref{fig:timeseries}a),  the clustering coefficient $C$ (Fig.~\ref{fig:timeseries}b) and the network adjacency correlation statistic $1-\gamma$ (Fig.~\ref{fig:timeseries}c) for each resulting snapshot time series $\vec{A}$ fir the seven days beginning on 11 October. The shortest rate $\Delta=5$ minutes exhibits high frequency noise overlaid on clear low frequency structure from the daily work cycle. As we would expect, as the snapshot rate increases, these high frequency variations are averaged out. When $\Delta=1440$ minutes (1 day), each day appears largely the same, with slight differences between week and weekend days; applying our analysis to the entire month of October shows that this regularity remains consistent over this longer period of time (data not shown).

\begin{figure} [t]%[htp]
\begin{center}
\includegraphics[scale=0.45]{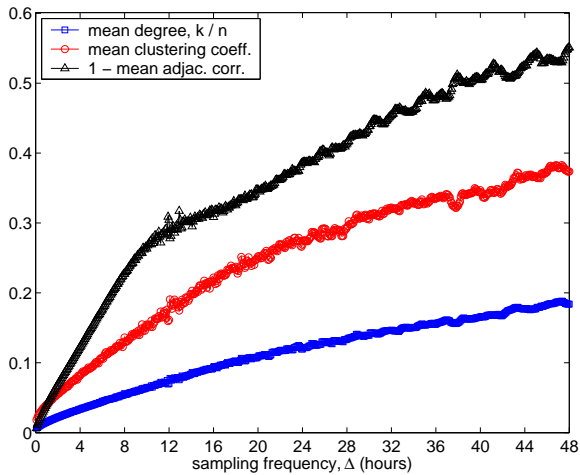}
\caption{The values of each network metric (mean degree $\langle k\rangle/n$, mean clustering coefficient $C$ and mean adjacency correlation $1-\gamma$) during the month of October as a function of snapshot rate $\Delta$; clearly, the value of each metric is proportional to the choice of $\Delta$. }
\label{fig:growth}
\end{center}
\end{figure}

As an aside, we consider the effect that using an increasingly large snapshot rate $\Delta$ has on the measured network statistics. Lengthening $\Delta$ has the effect of increasing the density of edges in each snapshot; thus, we would expect the mean degree and the clustering coefficient to increase with $\Delta$, while we would expect the network adjacency correlation to decrease. Figure~\ref{fig:growth} shows these statistics averaged over the set of snapshots derived from October, as a function of $\Delta$. The essentially monotonic growth of these curves illustrates that the choice of snapshot rate completely determines the measured value of the network statistics. Although the shape of these curves certainly conveys some information about how to choose $\Delta$, ultimately we would like a more principled way to make that choice.

\begin{figure} [t]%[hbp]
\begin{center}
\includegraphics[scale=0.45]{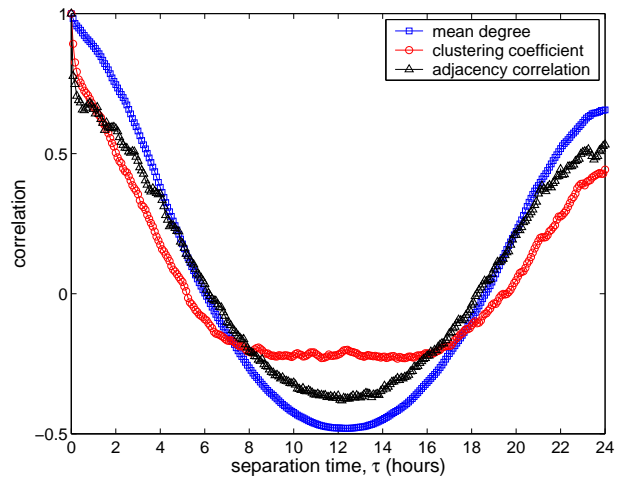}
\caption{The autocorrelation of three network metrics for $\Delta=5$ minutes during the week 11 October through 17 October. The correlations fall to zero at $\Delta=\{6.08, 5.25, 6.25\}$ hours respectively. }
\label{fig:autocorr}
\end{center}
\end{figure}

\begin{figure} [t]%[hbp]
\begin{center}
\includegraphics[scale=0.45]{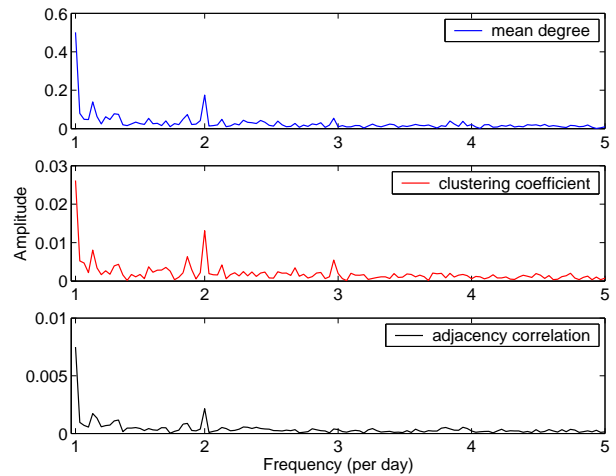}
\caption{The power spectra of three metric time series at $\Delta=5$ minutes over the course of the month of October. The principle peaks are at $\Delta=\{24,12\}$ hours, with additional minor peaks at $\Delta=8$ hours for the mean degree and clustering coefficient. }
\label{fig:power}
\end{center}
\end{figure}

Figure~\ref{fig:autocorr} shows the autocorrelation function of the time series in Figure~\ref{fig:timeseries} for $\Delta=5$ minutes, the smallest snapshot rate. This spectral analysis shows a strong -- and expected -- daily periodic behavior, but also that the autocorrelation for each statistic drops to zero at roughly the same time, being $\tau_{{\rm ACF}=0} = \{6.08, 5.25, 6.25\}$ hours, respectively. Figure~\ref{fig:power} shows the power spectra for the same time series, with strong peaks in both the mean degree and clustering coefficient spectra at roughly 1, 2 and 3 times a day, and at 1 and 2 times a day for the adjacency correlation. Thus, we suggest that the natural snapshot rate for dynamic proximity data is one half of the highest of these frequencies, or $\Delta_{\rm nat} = 4.08$ hours. Choosing this value gives a cross section of the curves on Figure~\ref{fig:growth}, and we report that the natural average degree $\langle k \rangle_{\rm nat} = 2.24$, the natural average clustering coefficient is $C_{\rm nat} = 0.084$ and the natural average adjacency correlation is $\gamma_{\rm nat} = 0.88$.

% ------ network metric growth as a function of \Delta
 
% ---------- Conclusions ----------
\vspace{3mm}\noindent {\bf \large Discussion}

Much of the work to-date on analyzing dynamic complex networks has focused on constructing a sequence of network snapshots, in which edges that may be varying rapidly in time are accumulated over a window of fixed length. In this manner, researchers have begun to move beyond static topology and consider questions related to how the behavior of processes changes when temporal ordering is taken into account~\cite{Kempe00} or how entities such as a social group or community may change over time~\cite{Saia06}. However, even the snapshot approach to representing inherently continuous topological variation poses its own set of problems, as the choice of snapshot rate $\Delta$ effectively determines many of the statistical properties of the resulting networks. As such, we make a cautionary note about the snapshot approach for studying dynamical social systems, as incorrectly choosing $\Delta$ may impose a strong bias on the resulting analysis and conclusions. To resolve this problem, here we suggest that there may be a natural snapshot rate for such dynamical systems that appropriately smoothes out high frequency variation while preserving important low frequency structural patterns.

% ------ autocorrelation of metrics
% ------

% ------ power spectra
% ------

The natural rate that we calculate for the dynamic proximity network $\Delta_{\rm nat}=4.08$ hours is presumably closely related to both the length of the day and the length of the human work day. Thus, the natural rate for other dynamical social systems may vary considerably, depending on the particular context of the social ties. In our case, this relatively long period may be related to the fact that the 66 core subjects of the MIT study worked in a modern office building; proximity networks in more physically active environments may exhibit a faster natural time scale.

Turning back to the larger question of studying dynamic social networks, the high frequency data of the MIT study allow us to observe that the network topology is itself evolving over a broad range of time scales. Although we have no clear distributional idealization of the data shown in Figure~\ref{fig:persistence}, it seems unlikely that they are power-law distributed~\cite{Clauset07}, or that they otherwise exhibit ``scale-free'' behavior. Rather, we suggest that proximity dynamics are a multi-scale phenomenon, with variation taking place at several distinct scales, likely driven by external periodicities of calendar cycles (e.g., day, week, month, season, year, etc.).

Finally, we suggest that the high frequency proximity network data could be used productively in network epidemiology by answering the question of what impact the temporal ordering of social contacts~\cite{Kempe00} has on the ultimate transmissibility of a disease, and thus the likely size of an epidemic. The advantage of this proximity data over more traditional approaches to network epidemiology is that it both removes any possibility for a self-reporting bias by directly and automatically observing human behavior, and directly records the network of who is close to whom, which is the network over which many (but not all) pathogens spread.

% ---------- Acknowledgements ----------
\vspace{5mm}
\noindent {\bf Acknowledgments.} This work was supported in part  by the National Science Foundation under grants PHY-0200909 and ITR-0324845, and by the Santa Fe Institute (AC).

% ---------- Bibliography ----------


\begin{thebibliography}{99}
\bibitem{footnote1}
Proximity is sampled periodically, and thus we do see some discretization on very short time-scales, as in the lower tail in Figure~\ref{fig:persistence}.
\bibitem{footnote2}
The clustering coefficient can be defined in two ways. Here, we use the convention that $C$ is the average of the number of triangles centered on a node $i$ divided by the number of paths of length two centered on node $i$.
\end{thebibliography}
\end{document}